\documentclass[11pt, a4paper]{article}

\usepackage[utf8]{inputenc}
\usepackage[english]{babel}
\usepackage{graphicx}
\usepackage{amsmath,amsthm}
\usepackage{amsfonts,amssymb}
\usepackage{natbib}
\usepackage{url,doi}
\usepackage{a4wide}
\usepackage{hyperref}			

\title{Electrical conductivity of the lowermost mantle explains absorption of core torsional waves at the equator}

\author{Nathanaël Schaeffer and Dominique Jault \\[0.5cm] \small
 Univ. Grenoble Alpes, CNRS, ISTerre, CS 40700, 38058 Grenoble Cedex 9, France
}

\begin{document}

\maketitle

\begin{abstract}
Torsional Alfv\'en waves propagating in the Earth's core have been inferred by inversion techniques applied to geomagnetic models.
They appear to propagate across the core but vanish at the equator, exchanging angular momentum between core and mantle.
Assuming axial symmetry, we find that an electrically conducting layer at the bottom of the mantle can lead to total absorption of torsional waves that reach the equator.
We show that the reflection coefficient depends on $G \tilde{B}_r$, where $\tilde{B}_r$ is the strength of the radial magnetic field at the equator, and $G$ the conductance of the lower mantle there.
With $\tilde{B}_r=7\times 10^{-4}$~T., torsional waves are completely absorbed when they hit the equator if $G\simeq1.3\times10^8$~S.
For larger or smaller $G$, reflection occurs.
As $G$ is increased above this critical value, there is less attenuation and more angular momentum exchange.
Our finding dissociates efficient core-mantle coupling from strong ohmic dissipation in the mantle.
\end{abstract}

\section{Introduction}

Geostrophic motions propagating outward from the cylindrical surface tangent to the inner core and vanishing, 3 to 4 years later, upon their arrival near the outer core equator have been detected  by \citet{gillet-et-al-2010}.
In order to estimate the magnetic field intensity  in the Earth's core interior, they relied on a model of one-dimensional torsional Alfv\' en waves propagating in a spherical shell filled by a perfectly conducting and inviscid fluid \citep{Braginsky:1970rm}. They found that the model reproduces well the propagation pattern of the geostrophic motions in a certain range of values of $G \tilde{B}_r^2$, where $G$ is the conductance of the mantle and $\tilde{B}_r$ is the  rms value of the radial field at the core-mantle boundary. Assuming that  $\tilde{B}_r$ is $7\times 10^{-4}$ T., they estimated the conductance to be between $6\times10^7$ S. and $2.8\times 10^8$ S.
They interpreted their results in terms of magnetic friction at the core-mantle boundary damping the torsional waves.
In this process, angular momentum is exchanged between core and mantle, contributing to the length-of-day variations.

\citet{gillet-et-al-2015} examined torsional wave propagation in the Earth's core over a longer time interval (1940-2010) and confirmed that the waves repeatedly  travel from the cylinder circumscribing the inner core to the outer core equator, where they disappear. Magnetic fields originating in the core are possibly delayed across the electrically conducting mantle but \citet{gillet-et-al-2015} also remarked that the phase retardation $\tau$ between changes in the length-of-day  and core angular momentum series calculated from magnetic data is very small, namely less than half a year (\citet{Holme:2013fk} argued for an even smaller phase lag $\tau\le 0.2$ yr).  Assuming spherical symmetry and that the conductivity varies smoothly with radius, \cite{Jault2015} inferred from this constraint on the low-frequency electromagnetic delay time of the mantle an upper limit on the mantle conductivity that corresponds to $G\le 3\times10^7$ S. This calculation leaves aside a possible thin layer of high conductance at the bottom of the mantle \citep{buffett2007} that may consist of post-perovskite \citep{ohta2008} or iron-rich (Mg,Fe)O assemblage \citep{wicks2010}.

Investigating torsional wave models from a theoretical standpoint, \citet{schaeffer2012} remarked that while the values of the viscosity and the magnetic diffusivity in the core interior are unimportant, their ratio $Pm$ (the magnetic Prandtl number) determines the reflection at the equator.
There, the core-mantle boundary can be approximated as a plane parallel to the rotation axis.
Using a one-dimensional Alfv\'en wave theory,  \citet{schaeffer2012} computed a reflection coefficient $R$, defined as the ratio of the outgoing fluid velocity amplitude over the incoming one:
\begin{equation}
R = \frac{1-\sqrt{Pm}}{1+\sqrt{Pm}}	,	\label{eq:R_insul}
\end{equation}
for an insulating wall.
With two-dimensional numerical simulations, they showed that the one-dimensional plane wave theory is a useful guide for torsional waves in the spherical geometry of the Earth's core, although there are some differences.
Hence, they argued that in the case of most geodynamo simulations, which have been performed with $Pm \sim 1$, there is no significant reflection of torsional Alfv\'en waves.
However, this effect can not explain the propagation pattern of geostrophic motions inferred by \citet{gillet-et-al-2010} as $Pm$ is about $10^{-6}$ in the Earth's core.
Here, we show that the value of $Pm$ becomes unimportant in the presence of an electrically conducting layer at the bottom of the mantle.
The reflection coefficient at the equator depends on $Q+\sqrt{Pm}$, where the dimensionless number $Q$ varies linearly with the conductance $G$ of the mantle. It is likely that $\sqrt{Pm}\ll Q$ in the geophysical situation.
We can thus interpret the propagation pattern of geostrophic motions in terms of poor reflection on the conducting mantle at the equator instead of magnetic damping distributed throughout the core.
Our estimate of the conductance $G$ that corresponds to poor reflection is about the same as the preferred value of \citet{gillet-et-al-2010}.

The paper is organized as follows.
We first combine, in section \ref{sec:ref-tw}, a theoretical study in planar geometry -- where a vertical wall models the core-mantle boundary next to the equator -- with a discussion of axisymmetric numerical simulations  in spherical geometry. We find that the reflection coefficient is correctly estimated from the study of a one-dimensional Alfv\' en wave hitting a conducting wall.
Section \ref{sec:discussion} is devoted to a discussion of the geophysical case. Energy dissipation in the equatorial region is a non monotonous function of the mantle electrical conductivity. Accordingly, torsional waves may propagate almost unhindered by magnetic friction in the core volume and yet be almost fully absorbed at the equator.

\section{Reflection of torsional waves at the outer core equator}
\label{sec:ref-tw}

\subsection{Plane wave theory}
\label{sec:plane-wave}

Using a plane wave approach, we can derive analytically the reflection of one-dimensional Alfv\' en waves propagating in the half-space $0<x$ on a conducting wall at $x=0$ (see the detailed calculation in Appendix \ref{sec:1DA}):
\begin{equation}
R \simeq \frac{1 - Q - \sqrt{Pm}}{1+ Q + \sqrt{Pm}}		\, ,		\label{eq:R_cond_Sinfty}
\end{equation}
with
\begin{equation}
Q =   \sqrt{\frac{\mu_0}{\rho}} \, G \, B \, , \qquad Pm=\frac{\nu}{\eta} \, ,
\label{eq:QW-first}
\end{equation}
$B$ the magnetic field perpendicular to the wall, $\rho$  the fluid density, $\nu$ its kinematic viscosity, $\eta$ its magnetic diffusivity, $\mu_0$  the magnetic permeability of free space and $G$  the conductance of the solid region ($-\delta\le x \le 0$):
\begin{equation}
G=\int_{-\delta}^0  \sigma(x) \mathrm{d}x \, .
\end{equation}
To obtain (\ref{eq:R_cond_Sinfty}), we have assumed that the electromagnetic skin depth in the solid domain is larger than the conducting layer thickness (low-frequency approximation), in which case $R$ is a real number.
In the limit of small $Pm$, which is relevant for liquid metals,  no reflection occurs when  $Q \simeq 1$.

\begin{figure}
\centering
\includegraphics[width=0.6\linewidth]{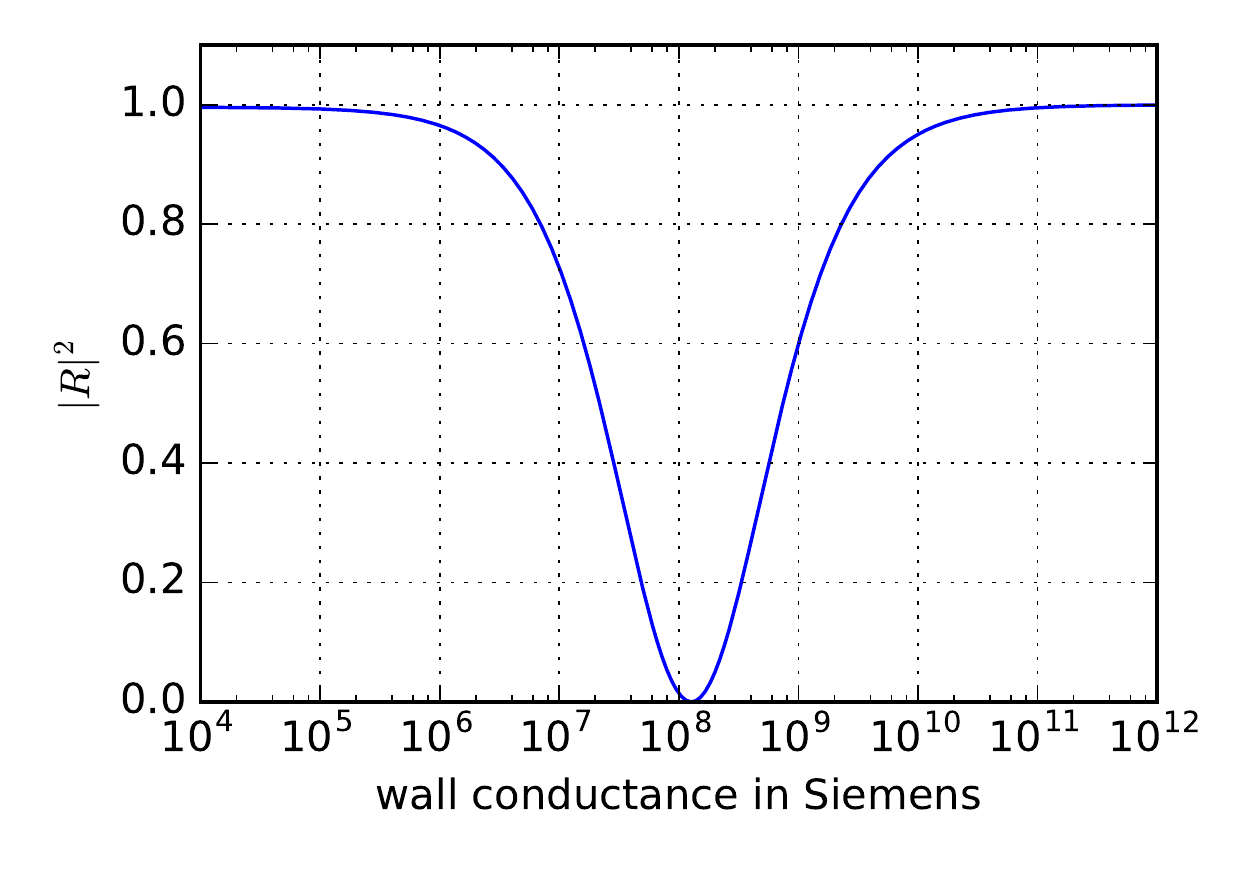}
\caption{Reflected energy of a one-dimensional Alfv\' en wave as a function of the conductance of the wall.
We set $Pm=10^{-6}$ and $V_A=6.24 \times 10^{-3}$~m/s corresponding to $B = 7 \times 10^{-4}$~T with $\rho=10^4$~kg/m$^3$, values which are representative of the Earth's core.}
\label{fig:R2_mantle_cond}
\end{figure}

Plotting $R$ as a function of $G$ for Earth-like values of $B = 7 \times 10^{-4}$~T, $\rho=10^4$~kg/m$^3$ and $Pm=10^{-6}$, we find that the reflected energy is close to zero for a solid conducting layer next to the fluid with a conductance of $G \simeq 10^{8}$~S., as seen in figure \ref{fig:R2_mantle_cond}.

The parameters $Pm$ and $Q$ determine where energy dissipation takes place (see Appendix \ref{app:dissip}). Whatever the value of $Pm$, there is always equal dissipation of magnetic and kinetic energy within the Hartmann boundary layer. The ratio of dissipation in the solid wall and in this fluid layer scales as $Q/2\sqrt{Pm}$ although the dissipation in the wall,
\begin{equation}
{Q(1+R)^2}=\frac{4Q}{(1+Q+\sqrt{Pm})^2} \, ,
\end{equation}
vanishes in the limit $Q\rightarrow \infty$. In this limit, which corresponds to $R=-1$, there is maximal exchange of momentum between the fluid and the solid at each reflection but no dissipation.
The value of $Q/\sqrt{Pm}$ establishes which of the Hartmann layer or the conducting solid wall plays the main part in the reflection mechanism (see also equation (\ref{eq:R_cond_Sinfty})).

\subsection{Simulations in spherical geometry}

We have performed numerical simulations of the propagation of a torsional Alfv\' en wave pulse in a spherical shell of inner radius $r_i=0.35$ and outer radius $r_o=1$
permeated by an axially symmetric magnetic field $\mathbf{B}$, which is the gradient of a potential.
The simulations are of the same type as the ones done by \citet{schaeffer2012}, except for the addition of a conducting solid layer of thickness $\delta$ surrounding the fluid.
The XSHELLS code is used to time-step the linearized Navier-Stokes equation coupled to the linearized induction equation:
\begin{eqnarray}
  \partial_t \mathbf{u} + 2\mathbf{\Omega} \times \mathbf{u} &=& -\nabla p^* + \nu \nabla^2 \mathbf{u} +\frac{1}{\rho\mu_0} (\nabla \times \mathbf{b}) \times \mathbf{B},\label{eq_u} \label{eq:ns} \\
  \partial_t \mathbf{b} &=& \nabla \times (\mathbf{u} \times \mathbf{B} - \eta \nabla \times \mathbf{b}), \label{eq_b}  \label{eq:ind}
\end{eqnarray}
where $\eta=(\mu_0\sigma)^{-1}$ is the magnetic diffusivity. The electrical conductivity $\sigma$ is uniform in the fluid and jumps to another uniform value $\sigma_m$ in the solid outer shell.

The imposed magnetic field is an external axial quadrupole $\mathbf{B}=B  \, (2z\mathbf{e_z}-s \mathbf{e_s})$.
In this context, we define the Alfv\' en velocity as $V_A=B/\sqrt{\mu_0 \rho}$.
We hold the following parameters fixed across all simulations:
the magnetic Prandtl number $Pm = 10^{-3}$, the Ekman number $E=\nu/r_0^2\Omega = 10^{-10}$, 
the Lehnert number $Le=V_A/\Omega r_o = 9.46 \times 10^{-4}$.
This implies a Lundquist number $V_A r_0/\eta \simeq 10^{4}$.
No-slip boundary conditions are used at $r=r_o$ while free-slip are used at $r=r_i$ to avoid a very thin Ekman layer there.
The inner-core is insulating, while a solid conducting layer of thickness $\delta = 0.025\, r_0$ is included at the bottom of the mantle.
In this region, we time-step the induction equation (\ref{eq:ind}) with $\mathbf{u}=0$.
The XSHELLS code uses the spherical harmonic transform of the SHTns library \citep{schaeffer2013} in the angular coordinates and finite differences in the radial direction. 
The finite difference scheme handles conductivity jumps as detailed by \citet{cabanes2014a}.
A semi-implicit time-stepping scheme is used with diffusive terms (for both the momentum and the induction equations) handled by a Crank-Nicolson scheme, while all other terms are treated explicitly by a second order Adams-Bashforth scheme.
In order to fully resolve the very thin boundary layers (both viscous and magnetic) that occur near $r=r_0$, we need $N_r=5000$ radial shells.
The spherical harmonic expansion, which is restricted here to axisymmetric functions, is truncated after harmonic degree $\ell_{max}=500$.

We have chosen the following initial condition which corresponds to an outgoing pulse (propagating towards the equator):
\begin{equation}
\mathbf{b} = \mathbf{u} = s \, \exp \left( -\frac{(s-s_0)^2}{d^2} \right) \mathbf{e_\phi},
\end{equation}
with $s_0 = (r_0+r_i)/2$ and either $d=0.02$ or $d=0.071$.

\begin{figure*}
\centering
\includegraphics[width=0.99\linewidth]{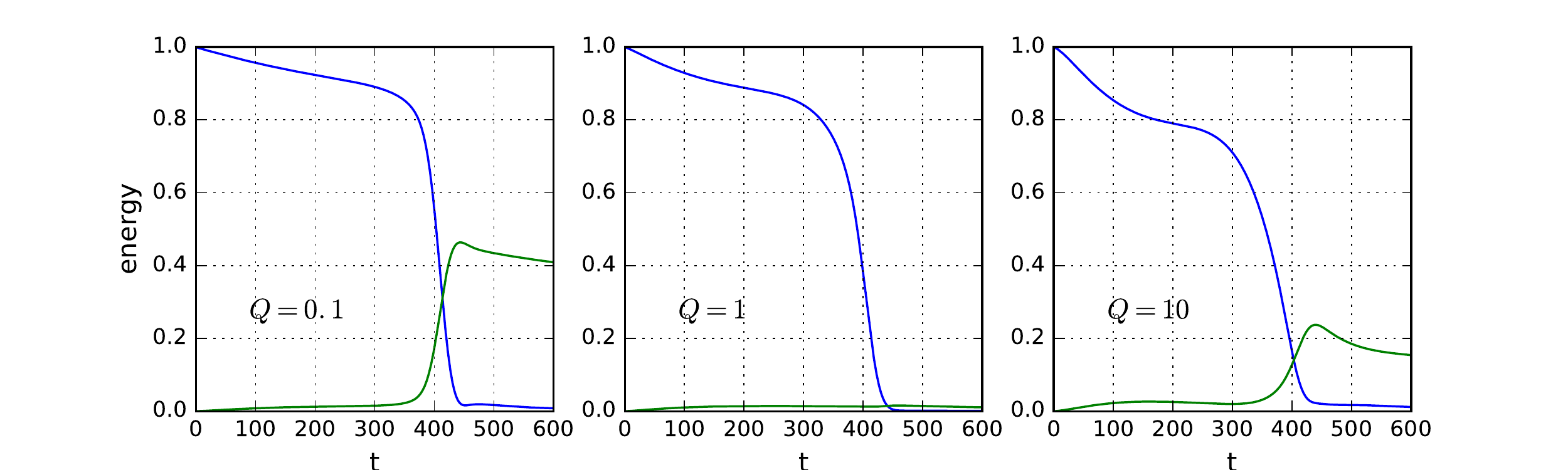}
\caption{Evolution of the energy of the incoming (blue) and reflected (green) pulses as a function of time in spherical shell simulations for $Pm=10^{-3}$, $E=10^{-10}$ and $S\simeq  10^{4}$.
Reflection occurs at about $t=400$.
There is no reflected pulse when the conductivity of the solid layer is such that $Q=1$.
}
\label{fig:Epm_vs_t}
\end{figure*}

\begin{figure}
\centering
\includegraphics[width=0.7\linewidth]{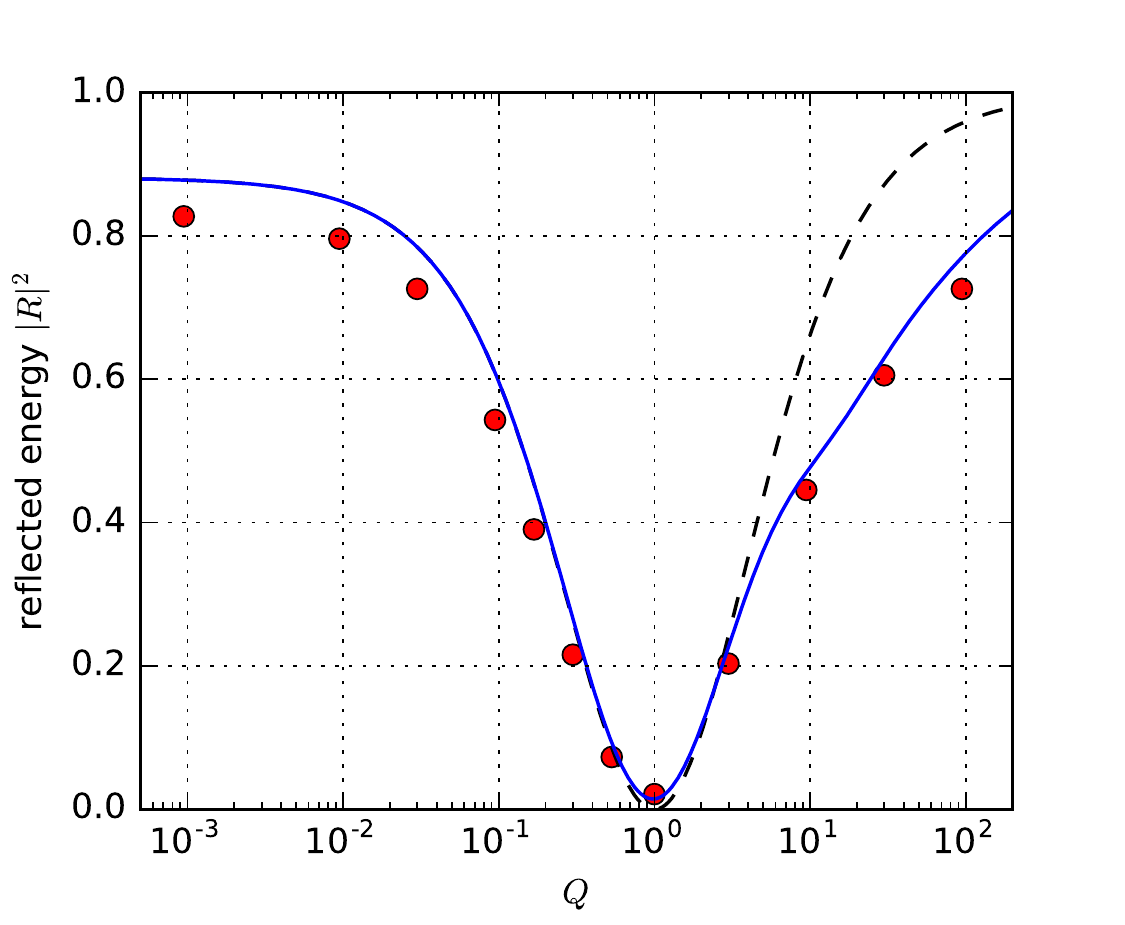}
\caption{Reflection of torsional Alfv\' en waves in spherical shell numerical simulations ($Pm=10^{-3}$, $E=10^{-10}$, $S \sim  10^4$).
Red dots are the measured reflected energy ratio $R^2$ in the simulations.
The blue line is the exact plane-wave theory (with $R$ depending on the wavelength), while the black dashed line is the thin layer approximation (which does not depend on the wavelength).
}
\label{fig:R_vs_Sm}
\end{figure}

We note $\eta_m = (\mu_0 \sigma_m)^{-1}$ the magnetic diffusivity in the solid conducting layer.
Twelve values of $\eta_m$, all larger than the magnetic diffusivity $\eta$ of the fluid, have been used, spanning five orders of magnitude.
We note
\begin{equation}
Q=  \sqrt{\frac{\mu_0}{\rho}} G B_r\vert_{r=r_o,\theta=\pi/2}, \qquad \mathrm{with} \qquad G=\int_{r_o}^{r_o+\delta} \sigma_m(r) \mathrm{d}r \, .
\end{equation}
Figure \ref{fig:Epm_vs_t} shows the evolution of the energy of the incoming and reflected wave fields for three of these simulations.
From this time history, we compute $|R|^2$, related to the reflection coefficient $R$.
It conforms with the theoretical prediction (\ref{eq:R_cond_Sinfty}) obtained from the study of one-dimensional Alfv\'en waves hitting a wall at $S \to \infty$. The agreement is especially good for $Q \sim 1$ as shown by figure \ref{fig:R_vs_Sm}.

%

\section{Discussion}
\label{sec:discussion}

In our study, we have used a simple magnetic field, with almost uniform strength and for parameters that have not been reached previously $E=\nu/r_0^2\Omega = 10^{-10}$ and $Pm = 10^{-3}$.
In addition, we have considered a uniform conductivity in the lowermost mantle together with a magnetic field independent of the longitude.

In this framework, attenuation of the torsional Alfv\'en waves that reach the equator is maximized in the vicinity of $Q=1$, which makes this value special for the study of torsional Alfv\'en modes.
However, we do not know yet whether $Q<1$, $Q=1$ or $Q>1$ at the Earth's core equator since the electrical conductivity of the lowermost mantle remains poorly known.
We find interesting to draw attention to the case $Q>1$, for which we obtain results that may first appear counter-intuitive.
For $Q=1$, the reflection coefficient at the equator changes sign. Accordingly, the waves reflected at $s=1$ have exactly zero amplitude for $Q=1$ while, as $Q$ is further increased above 1, there is more and more angular momentum deposited in the mantle although there is less and less ohmic dissipation.
The latter is not a monotonous function of the strength of the core-mantle magnetic coupling contrary to what had been previously accepted \citep{Dumberry:2008qv}.

Finally, we better understand how monitoring the geostrophic motions in the Earth's core together with the sub-decadal changes in the length of the day may constrain $Q$ at the core equator and, as a result, the electrical conductivity of the lowermost mantle.
$Q \simeq 1$ is compatible with conductance required by previous studies of electromagnetic core-mantle coupling \citep[e.g.][]{buffett2007,gillet-et-al-2010,roberts:2012} as well as with lowermost mantle composition \citep[e.g.][]{ohta2008, wicks2010}.
However, effectively constraining the conductivity of the lowermost mantle requires some additional work.
In particular, future studies should take into account the lateral heterogeneities of both magnetic field and mantle conductivity \citep[see e.g.][]{ohta2010,Ohta:2014db} that have been neglected in mantle electromagnetic filter theories and in this study of torsional wave absorption as well.

\section*{Appendix}
\appendix
\section{Reflection on a flat wall}		\label{sec:1DA}

We consider one-dimensional Alfv\' en waves, transverse to a uniform magnetic field, hitting a plane perpendicular to the imposed magnetic field  $B$ \citep{roberts1967}.
The field $B$ is directed along the $x$-axis, while the induced magnetic field $b(x,t)$ and the velocity field $u(x,t )$ are transverse to this field, along $y$.
The wall is placed at $x=0$.
Assuming invariance along the $y$ and $z$ axes, the problem reduces to a one-dimensional one, $u$ and $b$ depending only on $x$.
Projecting the Navier-Stokes equation and the induction equation on the $y$ direction (on which the pressure gradient and the non-linear terms do not contribute), we obtain the equations,
\begin{eqnarray}
\partial_t u &= & \frac{B}{\rho\mu_0} \partial_x b  + \nu \partial_{xx} u ,\label{eq:momentum} \\
\partial_t b &= & B \partial_x u + \eta \partial_{xx} b , \label{eq:induction}
\end{eqnarray}
in the half-space ($x>0$), where $\rho$ is density, $\mu_0$ is magnetic permeability of free space, $\eta$ is magnetic diffusivity and $\nu$ kinematic viscosity.
After scaling the magnetic fields to Alfv\' en speed units ($V_A = B/\sqrt{\rho\mu_0}$),
we transform (\ref{eq:momentum}-\ref{eq:induction}) into equations for the two Elsasser variables $h_\pm = u \pm b$.
Introducing a length scale $L$, the time-scale becomes $L/V_A$. The equations  of momentum and of magnetic induction combine into
\begin{equation}
\partial_t h_\pm \, \mp \partial_x h_\pm  - \frac{Pm+1}{2S} \partial_{xx} h_\pm = \frac{Pm-1}{2S} \partial_{xx} h_\mp,
\label{eq:prop-alf}
\end{equation}
 where  the Lundquist number $S$ is defined here as:
\begin{equation}
S  = \frac{V_A L}{\eta}	\, .
\end{equation}
The propagation of Alfv\' en waves requires that the dissipation in the bulk is small enough, which is ensured if $S \gg 1$ and $Pm \leq 1$.
We note that $h_-$ travels in the direction of the imposed magnetic field, while $h_+$ travels in the opposite direction.
Away from the boundary, the two variables $h_+$ and $h_-$ are independent.
At the boundary, reflection requires change of traveling direction, and thus transformation of $h_+$ into $h_-$.
If the wall is electrically insulating, and the fluid velocity vanishes on it (no-slip boundary condition),
we have $b=0$ and $u=0$, leading to $h_\pm = 0$.
These boundary conditions do not couple $h_+$ and $h_-$.
As a result, reflection is not allowed at an insulating and no-slip boundary when the coupling term on the right hand side of (\ref{eq:prop-alf}) vanishes ({\it i.e. }when $Pm=1$). 
For $Pm \neq 1$ the equations for $h_+$ and $h_-$ are coupled in the boundary layer.
This gives a mechanism for reflection of Alfv\' en waves at an insulating boundary \citep{schaeffer2012}.

We now insert a solid conducting layer of thickness $\delta$ and magnetic diffusivity $\eta_W$ between the solid insulator and the conducting fluid.
Equations \ref{eq:momentum} and \ref{eq:induction} in the fluid are complemented with
\begin{equation}
\partial_t b = \eta_W \partial_{xx} b \qquad \mbox{for}\qquad -\delta<x<0 \, . \label{eq:wall}
\end{equation}
At the insulator, we have $b(-\delta)=0$.
The magnetic field and the components of the electric field $\mathbf{E}$ parallel to the wall are continuous across the solid-liquid interface ($x=0$):
\begin{equation}
u(0)=0, \qquad \left. b \right|_{x=0^-} = \left. b \right|_{x=0^+} \, .
\end{equation}
Since $\mathbf{E}$  is related to the electric current $\mathbf{j} = j \mathbf{1_z} = \partial_x b\, \mathbf{1_z}$ by Ohm's law, and because $u(0)=0$, 
$j/\sigma$ is continuous across the boundary, which translates into
\begin{equation}
\left. \eta_W \, \partial_x b \right|_{x=0^-} = \left. \eta \, \partial_x b \right|_{x=0^+} \, .
\end{equation}

We seek solutions in the form of plane waves ${u,b} = {u_0, b_0} \exp\left( \mathrm{i}(\omega t + kx)\right)$ of frequency $\omega$ and wavenumber $k$.
In the fluid domain, we consider an interior solution and a boundary layer solution.
In the interior, the incoming and reflected waves have wavenumbers $\omega/V_A$ and $-\omega/V_A$, respectively. We also have $b=u$ for the incoming wave and $b=-u$ for the reflected one, which are characterized by the Elsasser variables $h^+$ and $h^-$, respectively.
We note $R$ the ratio of the outgoing velocity to the incoming one.
The solution in the fluid region matches the solution in the solid conducting region through a Hartmann boundary layer, where:
\begin{equation}
u(x)=u_{BL}\exp\left( -\frac{V_A}{\sqrt{\nu\eta}} x\right ), \quad b(x)=\sqrt{P_m} u(x) .
\end{equation}
In the conducting wall ($-\delta\le x \le 0$), we have
\begin{equation}
b(x)=c_1 \exp (\mathrm{i} k_W x) + c_2 \exp (-\mathrm{i} k_W x),
\end{equation}
with
\begin{equation}
k_W=\sqrt{\frac{\omega}{\eta_W}} \frac{1-\mathrm{i}}{\sqrt{2}} \, .
\end{equation}
Writing the matching condition at $x=0$ and the boundary condition at $x=-\delta$, we obtain a set of linear equations for $c_1$, $c_2$, $u_{BL}$,  and $R$.
In the limit $S^\dagger = V_A^2 / \omega\eta \gg 1$, we have
\begin{equation}
R = \frac{1 - Q^\dagger(\omega) - \sqrt{Pm}}{1+ Q^\dagger(\omega) + \sqrt{Pm}}	  \label{eq:R-general}
\end{equation}

with
\begin{equation}
Q^\dagger (\omega) = \frac{-\mathrm{i}V_A}{k_W\eta_W} F\left(\frac{\delta}{\lambda}\right), \quad
F(x) = \frac{1-\exp(-2x(1+ \mathrm{i}))}{1+\exp(-2x(1+ \mathrm{i}))}, \quad 
\lambda = \sqrt{\frac{2\eta_W}{\omega}} .
\label{eq:QW}
\end{equation}

When the thickness $\delta$ is small compared to the electromagnetic skin-depth $\lambda$, we can use the approximation $F(x \to 0) \sim x(1+\mathrm{i})$.
In this low-frequency limit, we have:
\begin{equation}
\lim_{\omega\rightarrow 0}	 (Q^\dagger(\omega))= Q	\, ,		
\end{equation}
where
\begin{equation}
Q = \frac{V_A\delta}{\eta_W} =  \mu_0V_A \,\sigma_W \delta \label{eq:def-Q}
\end{equation}
is a Lundquist number constructed with wall thickness and magnetic diffusivity. In this limit, (\ref{eq:R-general}) transforms into the expression (\ref{eq:R_cond_Sinfty}) of section \ref{sec:plane-wave}
and $R$ is a real number between $-1$ and $1$, which is independent of the frequency $\omega$.
When $\sigma_W=0$, we have $Q=0$ and we recover the reflection coefficient of the insulating wall \citep{schaeffer2012}.
Finally, it is straightforward to extend these results to the case of a thin conducting region consisting of a pile of $N$ layers ($1\le i \le N$)
of thickness $\delta^{i}$ and conductivity $\sigma_W^{i}$ with:
\begin{equation}
Q=\mu_0V_A \sum_{i=1}^N \sigma_W^{i} \delta^{i} \, .
\end{equation}

\section{Energy dissipation during reflection} \label{app:dissip}

Using $V_A/\omega$ as unit of length, the viscous dissipation $D_V$ in the Hartmann layer is:
\begin{equation}
D_V=\frac{Pm}{S^{\dagger}} \int_0^\infty (\partial_x u)^2 dx \, ,
\end{equation}
with
\begin{equation}
u(x)= -(1+R)\exp\left( -\frac{{V_A}^2}{\omega\sqrt{\nu\eta}} x\right ) \, .
\end{equation}
Thus, we have $D_V=\sqrt{P_m}(1+R)^2$ and the same result for the magnetic dissipation in the Hartmann layer.

We calculate the dissipation $D_W$ in the wall from the expression of the magnetic field in this solid region
\begin{equation}
b(x)=c_1 \exp (\mathrm{i} \frac{k_WV_A}{\omega} x) + c_2 \exp (-\mathrm{i} \frac{k_WV_A}{\omega} x),
\end{equation}
and the boundary conditions that have already been used to derive the reflection coefficient $R$ in Appendix \ref{sec:1DA}
\begin{equation}
-4 k_W^2 \delta^2 c_1^2 = Q^2 (1+R)^2 \, , \qquad
b(-\delta^{\dagger})  = 0 \, ,
\end{equation}
where $\delta^{\dagger}=\omega\delta/V_A$ is the dimensionless wall thickness.
In the low frequency limit $\delta \ll \lambda$, we have
\begin{equation}
D_W=\frac{1}{S_W^\dagger}\int_{-\delta^\dagger}^0 (\partial_x b)^2 dx =Q(1+R)^2 
\end{equation}
($S_W^\dagger=V_A^2/\omega\eta_W$).

\section*{Acknowledgments}

The authors wish to thank Nicolas Gillet and Elisabeth Canet for fruitful discussions, and two anonymous reviewers for their constructive comments.
The XSHELLS code used for the numerical simulations is freely available at \texttt{https://bitbucket.org/nschaeff/xshells}.
This work was partially supported by the French {\it Centre National d'\'Etudes Spatiales (CNES)} for the study of Earth's core dynamics in the context of the Swarm mission of {\it ESA}
and by the French {\it Agence Nationale de la Recherche} under the grant ANR-11-BS56-011.
Most of the computations were performed using the Froggy platform of the CIMENT infrastructure (\texttt{https://ciment.ujf-grenoble.fr}), supported by the Rh\^ one-Alpes region (GRANT CPER07\_13 CIRA), the OSUG@2020 labex (reference ANR10 LABX56) and the Equip@Meso project (reference ANR-10-EQPX-29-01).

\bibliography{cond-OT.bib}

\end{document}